\documentstyle[12pt]{aipproc}
\input epsf
\def\dofig#1{\vskip.2in\centerline{\epsfbox{#1}}}
\begin{document}
\hfill{FERMILAB-Conf-98/011-T}
\vskip .1cm
\hfill{January 1998}
\title{Compositeness Test at the FMC with Bhabha
Scattering~\footnote{talk given by S.~Keller at the {\it Workshop on
Physics at the First Muon Collider and the Front End of a Muon Collider}, 
Fermilab, November 6--9, 1997, to appear in the Proceedings}}
\author{E.~J.~Eichten and S.~Keller}
\address{
Fermi National Accelerator Laboratory, P.~O.~Box 500,\\ Batavia, IL
60510, U.S.A.}
\maketitle
\begin{abstract}
It is possible that quarks and/or leptons have substructure 
that will become manifest at high energies.
Here we investigate the limits on the muon compositeness scale that 
could be obtained at the First Muon Collider using Bhabha scattering.  
We study this limit as a function of the collider energy and
the angular cut imposed by the detector capability.
\end{abstract}

\section{Introduction}

The presence of three generations of quarks and leptons, apparently
identical except for mass, strongly suggests that they are composed of
still more fundamental fermions.
It is clear that, if substructure exist, the associated 
strong interaction energy scale $\Lambda $ must be
much greater than the quark and lepton masses. Long ago, ~'t~Hooft figured
out how interactions at high energy could produce essentially massless
composite fermions: the answer lies in unbroken chiral symmetries of the
underlying fermions and confinement of their new strong 
nonabelian gauge interactions\cite{thooft}.
There followed a great deal of theoretical effort to construct a realistic
model of composite quarks and leptons (see, e.g., Ref.~\cite{comp}) which,
while leading to valuable insights on chiral gauge theories, fell short
of its main goal.

It was pointed out that the existence of quark and lepton substructure
will be signalled at energies well below $\Lambda$ by the appearance
of four-fermion ``contact'' interactions which differ from those
arising in the standard model\cite{elp83,ehlq}.  These interactions
are induced by the exchange of bound states associated with the new
gauge interactions.  The main constraint on their form is that they
must be $SU(3) \otimes SU(2) \otimes U(1)$ invariant because they are
generated by forces operating at or above the electroweak scale. These
contact interactions are suppressed by $1/\Lambda^2$, but the coupling
parameter of the exchanges---analogous to the pion-nucleon and
rho-pion couplings---is not small.  Compared to the standard model,
contact interaction amplitudes are of relative order $s/(\alpha
\Lambda^2)$, where $\sqrt{s}$ is the center of mass energy of the
process taking place and $\alpha$ the coupling constant of the
standard model interaction.  The appearance of $1/\alpha$ and the
growth with $s$ make contact-interaction effects the lowest-energy
signal of quark and lepton substructure.  They are sought in jet
production at hadron and lepton colliders, Drell-Yan production of
high invariant mass lepton pairs, Bhabha scattering, $e^+e^-
\rightarrow \mu^+\mu^-$ and $\tau^+\tau^-$\cite{pdg}, atomic parity
violation\cite{rosner}, and polarized M\"{o}ller
scattering\cite{kumar}.  Hadron collider experiments can probe values of
$\Lambda$ from the 2--5~TeV range at the Tevatron to the 15--20~TeV
range at the LHC (See Refs.~\cite{ehlq,snow96}).

Here, we will study in some details one specific example for the First
Muon Collider (FMC): the constraint that can be imposed on the scale
of muon compositeness by measuring Bhabha scattering. The specific
form for the muon contact interaction is presented in Section II.
(All the results presented here are also applicable to electron
compositeness at $e^+e^-$ colliders with same energy and luminosity.)

CELLO at PETRA with a center of mass energy, $\sqrt{s}$, of 35 GeV and
an integrated luminosity, ${\cal L}$, of $86 \ {\rm pb}^{-1}$ was able
to put a lower limit on the (electron) compositeness scale of the
order of 2-4~TeV, depending on the specific model for
compositeness\cite{cello}.  This is about the same reach as the
current Tevatron reach.  This clearly show the potential for lepton
colliders to probe compositeness; they have an enormous reach.

In section III, we study the reach versus the energy of the FMC,
with the corresponding luminosity chosen for this workshop.  We
also study the effect of the angular cut on the reach.  It is
important to study that effect because large amount of radiation close
to the beam will limit the capability of the detectors outside the 
central region.

\section{Muon Compositeness}

We assume the muon has a substructure.  For collider energy below
the scale associated with this new structure, the effect can be parametrized 
by a four fermions interaction.  Here, we use the
flavor-diagonal, helicity-conserving contact interaction proposed by
E.~J.~Eichten, K.~Lane and M.~E.~Peskin\cite{elp83}:
\begin{equation}
\label{eq:contact}
{\cal L}= \frac{g^2}{2 \Lambda^2 }  \left[
\eta_{LL} \ j_L j_L
+ \eta_{RR} \ j_R j_R
+ \eta_{LR} \ j_L j_R \right],
\end{equation}
$j_L$ and $j_R$ are the left-handed and right-handed currents
and $\Lambda$ the compositeness scale.  The coupling constant,
$\alpha_{\rm new} \equiv \frac{g^2}{4 \pi} $, 
is assumed to be strong and set to one.
By convention, the $\eta $ have magnitude one.

The unpolarized cross section at lowest order, including the $\gamma$
and $Z$ exchange (s and t channel) and the contact interaction from
Eq.~\ref{eq:contact} can be written in the following form, see
Ref.~\cite{cello}:
\begin{equation}
\frac{d\sigma}{d\Omega}= \frac{\alpha^2}{8s} \left[ 4B_1+B_2(1-\cos \theta)^2 
+ B_3 (1+\cos \theta)^2 \right]
\end{equation}
where
\begin{equation}
B_1=\left( \frac{s}{t} \right)^2 \left| 1 + (g_V^2 - g_A^2) \ \xi +
\frac{\eta_{RL} t}{\alpha \Lambda^2} \right|^2 ,
\end{equation}
\begin{equation}
B_2=\left| 1 + (g_V^2 - g_A^2) \ \chi +
\frac{\eta_{RL} s}{\alpha \Lambda^2} \right|^2 ,
\end{equation}
\begin{eqnarray}
B_3&=&\frac{1}{2}\left| 1 + \frac{s}{t}+(g_V+g_A)^2 (\frac{s}{t}\xi+\chi) +
\frac{2 \eta_{LL} s}{\alpha \Lambda^2} \right|^2 \nonumber\\
& &\ \ \ \ \ +\frac{1}{2}\left| 1 + \frac{s}{t}+(g_V-g_A)^2 (\frac{s}{t}\xi+\chi) +
\frac{2 \eta_{RR} s}{\alpha \Lambda^2} \right|^2,
\end{eqnarray}
\begin{equation}
\chi = \frac{G_F}{2 \sqrt{2}} \frac{M_Z^2}{\pi \alpha}
\frac{s}{s-M_Z^2+i M_Z \Gamma},
\end{equation}
and
\begin{equation}
\xi = \frac{G_F}{2 \sqrt{2}} \frac{M_Z^2}{\pi \alpha}
 \frac{t}{t-M_Z^2+i M_Z \Gamma}.
\end{equation}
$\alpha$ is the usual fine structure constant, 
$\theta$ the scattering angle
between the incoming and outgoing muon, $t=-s/2 (1-\cos \theta )$,
$g_V$ and $g_A$ the vector and axial vector coupling constant, $M_Z$
and $\Gamma$ the mass and width of the Z, and $G_F$ the Fermi constant.
We will consider four typical models: LL couplings ( $\eta_{LL}=\pm
1$, $\eta_{RR}=\eta_{LR}=0$), RR couplings ( $\eta_{RR}=\pm 1$,
$\eta_{LL}=\eta_{LR}=0$), VV couplings (
$\eta_{LL}=\eta_{RR}=\eta_{LR}=\pm 1$), and AA couplings (
$\eta_{LL}=\eta_{RR}=-\eta_{LR}=\pm 1$).  The positive and negative
sign indicate the possible constructive or destructive interference
between the electroweak (EW) and compositeness contributions.
\begin{table}[b] 
\caption{Energy of the collider, luminosity, cross section ( with
$|\cos \theta | < 0.8$ ), and the expected number of events.}
\label{table1}
\begin{tabular}{|c|c|c|c|}
$\sqrt{s}$ (GeV) & ${\cal L}(fb^{-1})$ & $\sigma (pb)$ & N ($10^3$) \\ 
\hline
100        	& $.6$ 	& 125  			& 75 \\
200		& $1.$ 	& 34  			& 34 \\
350		& $3.$ 	& 11  			& 33 \\
500		& $7.$ 	& 5  			& 35 \\
\hline
\end{tabular}
\end{table}

\section{Experimental Bounds}

The total cross section for the different energy and luminosity
options considered at this workshop is presented in
Table~\ref{table1}.  The only detector effect included is an angular
cut: $|\cos \theta | < 0.8$.  No other detector effects were included
in this analysis.  As is well known, and can be seen in the set of
equations presented earlier, the EW cross section decreases
proportionally to $s$ (except in the Z resonance region), whereas the
interference term is independent of the energy and the pure
compositeness term increases with $s$.  This fact combined with the
(almost) constant number of events expected as a function of the
energy, see Table~\ref{table1}, clearly indicates that the best
compositeness limit will come from the highest energy option.

\begin{figure}
\epsfxsize=4.0in
\dofig{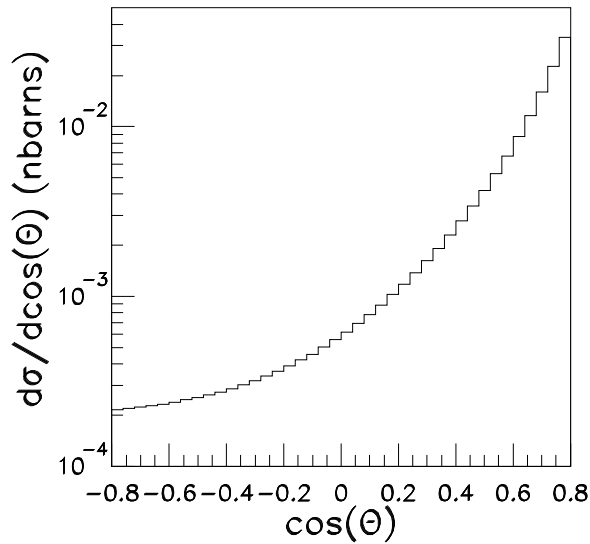}
\caption[angle]{$\cos \theta$ distribution at 500 GeV.}
\label{plot1}
\end{figure}

In Fig.~\ref{plot1}, the $\cos \theta$ distribution at $\sqrt{s} = 500$~GeV
is presented.  The typical t-channel, forward peaking is apparent.
The  $\cos \theta$ distributions at the other energies have the same shape
and are therefore not shown.

\begin{figure}
\epsfxsize=4.0in
\dofig{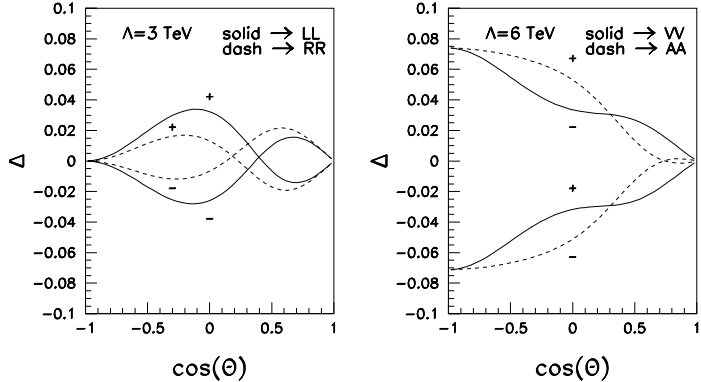}
\caption[models]{The variable $\Delta$ versus $\cos \theta$ at 
$\sqrt{s}=100$ GeV for the four models, LL, RR, VV, and AA, 
for the two signs of the $\eta$'s, indicate by $+$ and $-$ on the plot.}
\label{plot100}
\end{figure}

\begin{figure}
\epsfxsize=4.0in
\dofig{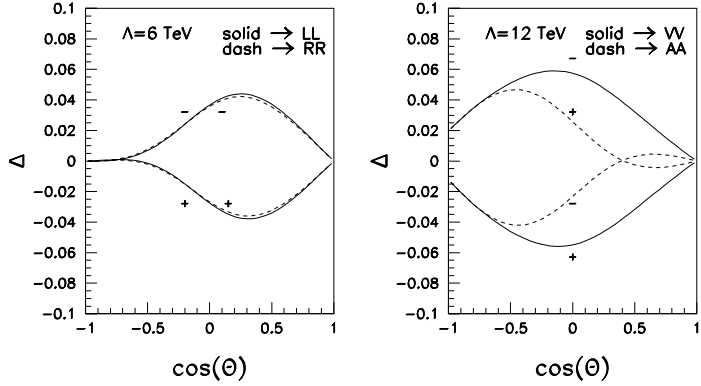}
\caption[LEPII]{Same as Fig. 2 at 200 GeV, about the LEPII energy.}
\label{plot200}
\end{figure}

\begin{figure}
\epsfxsize=4.0in
\dofig{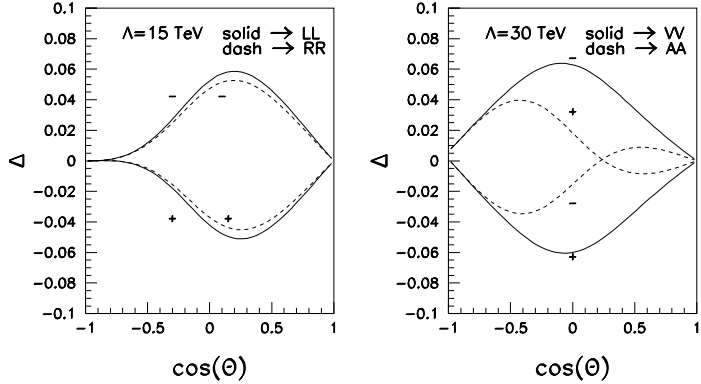}
\caption[FMC]{Same as Fig. 2 at 500 GeV.}
\label{plot500}
\end{figure}

To show the impact of the compositeness contribution we use the
variable $\Delta $, see Ref.~\cite{elp83}:
\begin{equation}
\Delta =
\frac{
\left ( \frac{d\sigma}{d\cos\theta} \right )_{EW+\Lambda}
- \left ( \frac{d\sigma}{d\cos\theta} \right )_{EW}  }
{\left ( \frac{d\sigma}{d\cos\theta} \right )_{EW}},
\end{equation}
the difference between the theory with and without the compositeness
terms, divided by the EW contribution.  The $\cos \theta$ distribution
of $\Delta$ is presented in Fig.~\ref{plot100} for $\sqrt{s}=100$~GeV.
The $\Lambda$'s were chosen such that the compositeness correction is
of the order of 10\% compared to the EW contribution.  That requires
that $\Lambda \simeq 30 \sqrt{s}$ for the LL or RR ($\frac{s}{\alpha
\Lambda^2} \sim .1$) couplings and $\Lambda \simeq 60 \sqrt{s}$ for
the VV or AA couplings ($\frac{s}{\alpha \Lambda^2} \sim .1/4$), there
are four interference terms in these latter models, see section~II.
The results for the four models are shown in Fig.~\ref{plot100} for
both sign of the $\eta's$.  It is clear that one can get limits on the
compositeness scale from the change of the shape of the distribution.
Note that in the forward region, in term of sensitivity to the
compositeness scale, the smaller change of the shape is compensated by
the larger number of events.  We also present the distribution at 200
and 500 GeV, in Fig.~\ref{plot200} and~\ref{plot500}, respectively.
The 300 GeV case is very similar to the 500 GeV case and is not shown.

The next step is to obtain a lower limit on $\Lambda$, assuming that
the data follow the EW theory.  Defining $x=1/\Lambda^2$, on average
(repeating the experiment many times) the central value of $x$
resulting from a $\chi^2$ fit will be zero because the data is assumed
to follow the EW theory.  For $x$ small enough, the differential cross
section is linear in $x$ (the $x^2$ term is small).  Within this
approximation the $\chi^2$ is quadratic in $x$, and the fit can be
trivially done.  The uncertainty on $x$, $\sigma_x$, is simply given
by two times the inverse of the second derivative of $\chi^2$ with respect
to $x$ (a constant).  We used
20 intervals for the fit, such that the lowest number of events in one
bin is still more than 100, which correspond to a maximum 10\% statistical
uncertainty in each bin.  The 95\% CL limit on $\Lambda$ is then
obtained from: $\Lambda^2 > 1 / (1.64 \sigma_x$).  The results are
presented in Table~\ref{table2}, for different energies of the muon
collider and $|\cos\theta|< 0.8$.
\begin{table}
\caption[LEP]{95\% CL limits (in TeV) for different energies (in GeV) of
the muon collider, we used $|\cos\theta|< 0.8$.  We also present the expected 
LEP limits for which we used $|\cos\theta|< 0.95$.}
\label{table2}
\begin{tabular}{|c|c|c|c|c|c|c|c|}
      & LEP(91) & LEP(175) & 100	& 200  & 350 & 500 & 4000 \\ \hline
${\cal L} (fb^{-1})$ & .15   & .1       & .6   & 1.    & 3.   & 7.   & 450. \\ \hline
LL    &	4.0  & 5.8       & 4.8  & 10   & 20  & 29  & 243 \\
RR    & 3.8  & 5.7 	 & 4.9  & 10   & 19  & 28  & 228 \\
VV    & 6.9  & 12. 	 & 12   & 21   & 36  & 54  & 435 \\
AA    & 3.8  & 7.2 	 & 12   & 13   & 21  & 32  & 263 \\
\hline
\end{tabular}
\end{table}
As expected the highest energy machine put the strongest constraint on
the compositeness scale.  The limit that the 4~TeV machine will be
able to put (with the luminosity scaled to maintain the number of
events constant) is really impressive.  The $\Lambda$ limits are large
enough such that the approximation used ($x$ small) is valid.  Because
of the approximation used, central value of $x$ equal to zero and the
differential cross section linear in $x$, the limits are independent
of the sign of the $\eta$.  We have only included the statistical
uncertainty in this analysis, and the limits presented here should be
considered within that context.  In particular, the absolute
normalization, which is used in this analysis, might be subject to
large uncertainties.  Our calculated limits for the CELLO case are
compatible with their measurements (their central $x$ value is of
course non zero).  In Table~\ref{table2}, we also have included the
expected limit for LEP with the current integrated luminosity (per
experiment) and its larger $\cos \theta$ coverage.

In Table~\ref{table3} we explore the effect of the $\cos\theta$ cut on
the 95\% CL limit.
Although any increase in coverage obviously
increases the limit, the improvement between $0.8$ and $0.95$ is less than
$10\%$.  We therefore conclude that the coverage up to $0.8$ is adequate
for this measurement.  It is not necessary to go down very close to
the beam to do a very good measurement.
\begin{table}
\caption[cuts]{95\% CL limits (in TeV) for different angular cuts at 
$\sqrt{s}=500$~GeV,
${\cal L}=7 fb^{-1}$.}
\label{table3}
\begin{tabular}{|c|c|c|c|c|}
$|\cos \theta| < $ & .6&.8&.9&.95 \\ \hline
LL&26&29&31&32\\
RR&24&28&30&30\\
VV&50&54&56&57\\
AA&28&32&34&35\\
\end{tabular}
\end{table}

\section{Conclusion}

We investigated the limits on the muon compositeness scale that could
be obtained at the First Muon Collider using the Bhabha scattering
process.  We considered four typical models for the four-fermion
contact terms expected as a low-energy signal for compositeness: LL,
RR, VV, and AA couplings.

As expected, the reach increases rapidly with energy.
We find that the reach at the 500 GeV FMC is $\Lambda > 30-55$~TeV depending 
on the model.  
At a future $4 \ {\rm TeV}$ muon collider the range extends to   
$\Lambda > 230-440$~TeV.  

The likelihood of limited angular coverage in detectors (because of
the unavoidable background of decaying muons) does not appear to poise
a severe problem for the study of muon compositeness.  We found that
an angular coverage corresponding to $| \cos \theta | < 0.8$ is
adequate to obtain 90\% of the full reach in the compositeness scale
$\Lambda$.

A number of detailed studies remain to be done.  For example, it is
clear that the polarization will help to differentiate between the
four models considered here.  Also we have only considered the
statistical uncertainties, the systematic uncertainties in
measurements of Bhabha scattering could be significant and need to be
included in future more realistic studies.

\end{document}